\documentclass[twocolumn,nopacs,preprintnumbers]{revtex4}
\usepackage{graphicx}
\usepackage{bm}
\usepackage{color}
\usepackage[normalem]{ulem}

\usepackage{physics}
\usepackage{subfigure}

\graphicspath{%
    {converted_graphics/}
    {/}
}
\begin{document}

\title{Decoherence and collective effects in critical media}


\author{F. M. D'Angelis}
\affiliation{Instituto de F\'isica, Universidade Federal do Rio de Janeiro, Caixa Postal 68528, Rio de Janeiro 21941-972, RJ, Brazil}
\author{F. A. Pinheiro}
\affiliation{Instituto de F\'isica, Universidade Federal do Rio de Janeiro, Caixa Postal 68528, Rio de Janeiro 21941-972, RJ, Brazil}
\author{F. Impens}
\affiliation{Instituto de F\'isica, Universidade Federal do Rio de Janeiro, Caixa Postal 68528, Rio de Janeiro 21941-972, RJ, Brazil}

\date{\today}

\begin{abstract}
We investigate the influence of critical phenomena on two distinct physical processes: the center-of-mass decoherence of a single emitter, and the collective radiation of two emitters.   We  address these different physical mechanisms with an unified formalism relying on standard  perturbation theory. We decompose the decoherence and the collective emission rates as a sum of two contributions, accounting for the spontaneous emission and for interference effects respectively. The former is enhanced by the Purcell effect when the material is in the vicinity of the emitter(s). The latter, associated to quantum interferences, experiences a ``sudden death''  near the critical point of the phase transition. Our findings unveil the interplay between the Purcell and collective effects and its dependence on metal-insulator transitions. We discuss two specific examples of phase transitions: the percolation transition in a metal-dielectric composite, and the metal-insulator transition in $\rm{VO}_2$. In the latter, decoherence and collective emission rates exhibit a characteristic hysteresis that strongly depends on the material temperature. These results, based on experimental data, suggest that $\rm{VO}_{2}$ could be explored as a versatile material platform where decoherence and collective emission can be tuned by varying the temperature.

\end{abstract}

\maketitle

\section{INTRODUCTION}

The study of collective effects has a long history in the context of spontaneous emission from an ensemble of quantum emitters (atoms, molecules or quantum dots) in both optical cavities and free space. The pioneering work of Dicke~\cite{dicke} has shown that  the intensity radiated by $N$ atoms placed at subwavelength distances may scale as $N^2$ for specific global quantum states of the emitters. This phenomenon, known as superradiance, arises from constructive interferences in the collective emission process. More recently, progresses in the field of nanophotonics have motivated and enabled the experimental investigation of superradiance in a broad range of photonic environments. Relevant examples are the coupling of quantum emitters (atoms, molecules, and quantum dots) to plasmonic waveguides~\cite{cano2010}, microcavities~\cite{temnov2005,pan2010}, metallic interfaces~\cite{choquette2010}, and epsilon-near-zero metameterials~\cite{fleury2013}. Tailoring and tuning collective optical effects enables one to enhance on demand light-matter interactions at the nanoscale. In order to develop novel photonic applications where collective spontaneous emission rate of several quantum emitters can be tuned, material platforms such as plasmonic waveguides~\cite{li2016}, graphene-based structures~\cite{huidobro2012} and nanofibers~\cite{solano2017} have been considered.

Collective effects, combined with a suitable photonic environment, have also interesting applications in the field of quantum control. Collective effects near a cavity have been used to put two ions in a maximally entangled state~\cite{casabone2015}. Previous work have considered using subradiance~\cite{pavolini1985,kaiser2012,scheibner2007,scully2015,guerin2016} -- the counterpart of superradiance associated to an inhibition of radiation through destructive interferences --  for quantum memories~\cite{kalachev2006}, nanolasers~\cite{leymann2015}, and quantum computers~\cite{kalachev2007}.

On the other hand, a key element to achieve a reliable quantum control is the design of a quantum environment minimizing decoherence~\cite{Kurizki2008}. The design of such an environment is a prerequisite when considering quantum information processing architectures at the nano-scale, where fluctuation-induced phenomena may predominate and compete with coherent processes. In this line, it has been shown that Casimir-Polder forces may be strongly affected by collective effects~\cite{sinha2018}, resulting from an interplay~ between the Purcell~\cite{Purcell46} and the Dicke effects~\cite{Fuchs18}. This approach suggests that collective effects and fluctuation-induced phenomena should be apprehended together.

In this article, we put forward a strategy that combines an alternative material platform, composite media and vanadium dioxide (${\rm VO}_2$), and a novel physical mechanism, critical phenomena, to achieve a control of either collective effects of several emitters or center-of-mass decoherence of a single emitter. There are evidence, both theoretical~\cite{juanjo1,juanjo2,Szilard16} and experimental~\cite{carminati}, that phase transitions affect spontaneous emission of a single emitter in a crucial way. This has been recently confirmed theoretically, setting the grounds for the determination of critical exponents via the Purcell factor~\cite{mbsn}. However, the effect of phase transitions in the collective emission or in the center-of-mass decoherence of quantum emitters remains unexplored so far. 

Here, we fill this gap by investigating the spontaneous emission of two-level quantum emitters and the center-of-mass decoherence of a single emitter near a composite medium that undergoes a metal-insulator phase transition. We use a formalism developed in the context of surface-induced dynamical Casimir phases~\cite{Francois13a,Francois13b,Francois14}, which enables an analogy between collective effects and the decoherence rate of a delocalized emitter~\cite{Reinaldo16}. We obtain a description of the center-of-mass decoherence rate as a function of the order parameter of the material phase transition. In particular, we demonstrate the strong suppression of the interference contribution to both the decoherence and to the collective emission rates in the presence of a media undergoing a percolation phase transition. We also investigate the specific example of the percolation transition in a $\rm{VO}_{2}$ material, to show that the decoherence and collective emission rates can exhibit a characteristic temperature-dependent hysteresis. Altogether our findings not only unveil the role of critical phenomena on collective, coherent interactions in quantum processes involving decoherence and collective emission, but they also suggests that $\rm{VO}_{2}$ could be explored as an alternative material platform to control these processes via thermal effects.

\color{black}

\section{METHODOLOGY}

In this Section, we describe the theoretical tools that will be used to address simultaneously two distinct physical phenomena in the vicinity of a medium undergoing a phase transition: the decoherence of a single emitter in a quantum superposition of two wave-packets, and the spontaneous emission of two quantum emitters. In the following we show that these two distinct physical phenomena can be described by the same theoretical framework, which enables us to unveil the role of interference effects in both situations.

\subsection{Interaction of the emitter with the light field}

The interaction of the emitter with the light field is described by a dipolar Hamiltonian $H(\vb{r},t) = - \mathbf{d}(t) \vdot \vb{E}(\vb{r},t),$ involving  the dipole operator  $\mathbf{d}(t)$ and the quantized electric field operator $\vb{E}(\vb{r},t)$  taken at the position $\vb{r}$ of the quantum emitter~\cite{Footnote1}. We work from now on in the interaction picture, letting the dipole and light field operators evolve according to the free dipolar and light field Hamiltonians. To obtain quantitative estimates of the decoherence rates, we have take this emitter as a two-level atom with transition wavelength $\lambda_0 = 450 \: {\rm \mu m}$.

\subsection{Electric field Green's function}

\begin{figure}[h!]
    \centering
    \includegraphics[width=8.5cm]{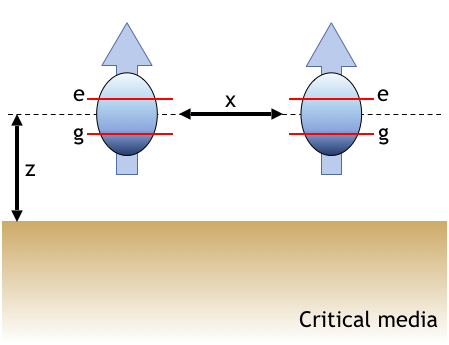}
    \caption{\label{fig:Boneco1}(color online) Schematic view of the systems under consideration, showing a representation of the two atoms or wave-packets for a two-level quantum emitter, separated by a distance $x$ from each other and at a distance $z$ to the material. In the case of two emitters we consider both to be with the dipoles aligned in the $z$ direction, while for the single emitter in a superposition of wave-packets we average over orientations.}
\end{figure}

The Green's dyadic in the vicinity of a bulk material contains a free part and a scattered part~\cite{NovotnyBook}
\begin{equation}\label{eq:G}
    G(\vb{r}_1,\vb{r}_2;\omega_0) = G^0(\vb{r}_1,\vb{r}_2;\omega_0) + G^{\rm Sca}(\vb{r}_1,\vb{r}_2;\omega_0)
\end{equation}
where $G^0(\vb{r}_1,\vb{r}_2;\omega_0)$ and $G^{\rm Sca}(\vb{r}_1,\vb{r}_2;\omega_0)$ are the free and scattering Green's dyadic, respectively. We shall consider situations where the two wave-packets (or two quantum emitters)  are located at the same height $z$ with respect to the material surface, and at a distance $x$ as depicted on Fig.~\ref{fig:Boneco1}. We shall also use the free and scattering contributions to the Green's function trace, denoted by $\mathcal{G}^{\rm 0}(\vb{r}_1,\vb{r}_2;\omega_0)$ and $\mathcal{G}^{\rm Sca}(\vb{r}_1,\vb{r}_2;\omega_0)$, respectively: 
\begin{eqnarray}\label{eq:TrG0}
   \mathcal{G}^{0}(\vb{r}_1,\vb{r}_2;\omega_0) & = & \frac{i}{2\pi} \int_0^\infty \frac{k_\parallel}{k_z} J_0(k_\parallel  x ) \dd k_\parallel \\
\label{eq:TrGSca}
    \mathcal{G}^{\rm Sca}(\vb{r}_1,\vb{r}_2;\omega_0) & = & \frac{i}{4\pi} \int_0^\infty \frac{k_\parallel}{k_z} J_0(k_\parallel  x)[r^{\rm TE,TE}   \\ & & + \frac{c^2}{\omega_0^2} r^{\rm TM,TM} (k_\parallel^2 - k_z^2)] e^{2i k_z z} \dd k_\parallel \nonumber
\end{eqnarray}
where $k_z = \sqrt{\frac{\omega_0^2}{c^2} - k_\parallel^2}$, $\omega_0$ is the transition frequency, $J_0$ is the zero-order spherical Bessel function, and $r^{\rm TE,TE}$ ($r^{\rm TM,TM}$) is the Fresnel reflection coefficient for TE (TM) polarization.

\subsection{Modeling the optical properties of the materials}\label{subsection:opticalproperties}

We model the optical properties of the semi-infinite, nonmagnetic critical media via the Bruggeman Effective Medium Theory (BEMT)~\cite{Bruggeman35,PercolationBook}, which is one of the simplest analytical models that can predict the percolation phase transition. This model is well suited for the two phase transitions considered below, namely the percolation transition in a metal-dielectric composite material and the metal-insulator transition in $\rm{VO}_2$~\cite{Choi96}. The system considered for the percolation transition is a polystyrene host-medium filled with gold inclusions. In the $\rm{VO}_2$ material, metallic clusters are formed as the temperature increases, so that they can also be treated as effective metallic inclusions. ${\rm VO}_2$ is particularly interesting for photonic and electronic applications because its transition occurs near room temperature, allowing for novel applications such as metamaterial reflectors and switches~\cite{muskens, markov}.  This enables one to address both phase transitions with the same effective model theory, where the relevant critical parameter is the filling factor $f$ representing the fraction of the volume occupied by these inclusions. The effective permittivity $\epsilon_{\rm eff}$
is the solution with positive imaginary part (as expected for a passive medium) of the following equation~\cite{Bruggeman35,PercolationBook}
\begin{align}\label{eq:epsilon_eff}
   & (1-f)\left( \frac{\epsilon_{\rm hm} - \epsilon_{\rm eff}}{\epsilon_{\rm eff} + L(\epsilon_{\rm hm} - \epsilon_{\rm eff})} + 
    \frac{4(\epsilon_{\rm hm} - \epsilon_{\rm eff})}{2\epsilon_{\rm eff} + (1-L)(\epsilon_{\rm hm} - \epsilon_{\rm eff})}  \right) \nonumber \\
   & + f \left( \frac{\epsilon_{\rm i} - \epsilon_{\rm eff}}{\epsilon_{\rm eff} + L(\epsilon_{\rm i} - \epsilon_{\rm eff})} + 
    \frac{4(\epsilon_{\rm i} - \epsilon_{\rm eff})}{2\epsilon_{\rm eff} + (1-L)(\epsilon_{\rm i} -c)}  \right) = 0 \,,
\end{align}
where $\epsilon_{\rm hm}$ and $\epsilon_{\rm i}$ are the host medium and inclusion permittivities, respectively, and $L$ is the depolarization factor related to the geometry of the inclusions. In the percolation transition, the filling factor $f$ and the geometry of the inclusions can be chosen independently from the temperature. Here we consider spherical inclusions corresponding to a depolarization factor $L = 1/3$. Differently, for $\rm{VO}_2$, the depolarization factor $L$, the filling fraction $f$ and the insulator (metallic) permittivities $\epsilon_{\rm hm}$ ($\epsilon_{\rm i}$) are determined by the temperature $T$. In the latter, we have used a relation between the filling fraction $f$ and the depolarization factor $L$ based on the experimental data reported in Ref.~\cite{VO2ExpDataPRB09}. For both materials, the phase transition occurs at the critical value of the filling factor $f = 1/3,$ for which the effective permittivity $\epsilon_{\rm eff}$ becomes purely imaginary~\cite{Bruggeman35,PercolationBook}.  

\subsubsection{Metal-dielectric composite undergoing a percolation transition}

In order to describe the material parameters, we use the Drude model for the inclusions and the Drude-Lorentz model for the host medium, with parameters extracted from experiments. According to these models, the permittivity is expressed as $\epsilon_{\rm i}(\omega)=1- \omega_{\rm p \: i}^2/(\omega^2+ i \gamma_{\rm i} \omega)$ for the gold inclusions and as $\epsilon_{\rm hm}(\omega)=1+ \sum_j \omega^{\rm hm \: 2}_{ {\rm p} j}/(\omega_{R j}^2-\omega^2 - i \Gamma_j \omega)$ for the host medium. We have introduced the plasma frequencies for the inclusions $\omega_{\rm p \: i}$ and the oscillating strengths $\omega^{\rm hm}_{ {\rm p} j}$ for the host medium, as well as the inverse of the relaxation times $\gamma_{\rm i}$ and $\Gamma_{\rm j}$ for the inclusions and for the host medium, respectively and the resonant frequencies $\omega_{R j}$. The experimental values of these parameters were extracted from Ref.~\cite{Hough80,Ordal85}.

\subsubsection{$\rm{VO}_2$ material undergoing a metal-insulator phase transition}

In contrast to the metal-dielectric composite, the permittivities of the $\rm{VO}_2$ material now rely completely on real experimental data taken from a thin film of $200 \rm{nm}$ thickness and deposited over a $1 \rm{mm}$ thick sapphire ($\rm{Al}_2\rm{O}_3$) substrate~\cite{VO2ExpDataPRB09}. As a manifestation of the material hysteresis, there is a discrepancy in the temperatures associated to the phase transition in the cooling and heating cycles. Indeed, the phase transition occurs at the temperature $T \approx 342$K for the heating cycle and at the temperature $T \approx 336$K for the cooling cycle~\cite{VO2ExpDataPRB09}.

\section{CENTER-OF-MASS DECOHERENCE OF A SINGLE EMITTER NEAR A CRITICAL MEDIUM}

In this Section, we investigate the decoherence rate suffered by a delocalized single quantum emitter nearby a material undergoing a phase transition. Specifically, we consider a neutral two-level atom with no permanent dipole, but most of the discussion could be extended to other quantum emitters. The decoherence rate involves two contributions~\cite{Reinaldo16}, corresponding to a first term simply associated to the global photon emission rate, and a second term involving quantum interferences and related to the quality of the which-way information.  We study the interplay between these two contributions as the material undergoes a phase transition. As demonstrated below, near the critical point of the transition, the contribution associated to interference effects is drastically suppressed at the critical point.

\subsection{Problem statement and general discussion}

As in Ref.~\cite{Reinaldo16}, we assume that a single quantum emitter is initially in an excited state, and that its center-of-mass wave-function involves a superposition of two well-separated wavepackets with negligible overlap. We monitor here the external degrees of freedom of this quantum emitter (the considered quantum system), which may produce at most a single photon. The internal degrees of freedom (d.o.f.) of the emitter, the light field and 
 the material d.o.f. play the role of an environment surrounding the quantum system. The phase transition affects the conductivity properties of the material and thus influences this environment. By virtue of the Purcell effect~\cite{Purcell46}, the presence of the material near the emitter significantly increases the photonic emission.  This phenomenon has been shown to be greatly enhanced near the critical point of the phase transition~\cite{Szilard16}. As the emitted photon may reveal the position of the emitting wavepacket, the decoherence rate suffered by the wavefunction of the quantum emitter is also drastically increased near the critical point.

The discussion to follow focuses, however, on a more subtle influence on decoherence. Beyond the rate of photonic emission, the decoherence rate also depends on the quality of the which-way information contained in the emitted photon. The amount of which-way information depends in turn on intrinsic features of light propagation, as well as on the distance between the wavepackets composing the quantum emitter wavefunction. For instance, the emission of a long wavelength (larger than the wavepacket separation) photon would hardly affect the wavefunction of the emitter, as an observer could not reliably infer from a field measurement the emitting wavepacket. In contrast, the emission of a short wavelength photon is expected to destroy the wavefunction coherence, as an observer could in principle identify unambiguously the emitting wavepacket.  As the production of a long wavelength photon carries a poor quality which-way information, it is associated to a negative interference contribution to the decoherence rate (``recoherence'' rate) which almost cancels the contribution proportional to the photonic emission~\cite{Reinaldo16}.

More generally, the structure of the electric field Green's function affects the amount of which-way information per emitted photon.  We put in evidence here the role played by the interplay between the direct light propagation and the light propagation after scattering by the nearby material.  As the material undergoes a phase transition, the scattering contribution to the electric field Green's function exhibits a sharp variation, which strongly reduces 
proportionally the interference contribution to the decoherence rate. 
 
\subsection{General expressions of the local and non-local decoherence rates}

\label{section:decoherence}

For convenience, we first recall the definition of the non-local and local decoherence rates~\cite{Reinaldo16}. We take the initial quantum state of the two-level atom as a product state $| \Psi(0) \rangle = | e \rangle \otimes \frac {1} {\sqrt{2}} \left( | \psi_+ \rangle + e^{i \theta_0} | \psi_- \rangle \right),$ and assume the light field to be initially in the vacuum state.  $| \psi_{\pm} \rangle$ denote external wavepackets of size assumed to be much smaller than the  transition wavelength.
The local and non-local decoherence rates can be obtained by considering the corresponding imaginary phases in the long-time limit~\cite{Reinaldo16}:

\begin{equation}\label{eq:dec_rate_total}
    \Gamma^{\rm Dec}_{\rm L, NL} = \lim_{\Delta t \to \infty} \frac{1}{ \Delta t} \Im( \varphi_{\rm L,NL}(\Delta t)  )  
\end{equation}

The complex local $\varphi_{\rm L}(\Delta t)$ and non-local $\varphi_{\rm NL}(\Delta t)$ phases arise from the dipolar interaction during the duration $\Delta t,$ and are given by standard time-dependent perturbation theory as

\begin{eqnarray}\label{eq:S_I}
    \varphi_{\rm L}(\Delta t) &= \displaystyle  \frac{i}{2\hbar^2} \sum_{k=1,2} \int_0^{\Delta t} \!\!  \dd \tau \int_0^{\Delta t} \dd \tau' \expval{H(\vb{r}_k,\tau)H(\vb{r}_k,\tau')} \nonumber \\
     \varphi_{\rm NL}(\Delta t) &= \displaystyle -\frac{i}{\hbar^2} \int_0^{\Delta t} \dd \tau \int_0^{\Delta t} \dd \tau' \expval{H(\vb{r}_1,\tau)H(\vb{r}_2,\tau')} 
     \label{eq:localnonlocal phases}  
\end{eqnarray}

The local phase contains a sum of two contributions involving a single wavepacket, while the non-local phase cannot be split in such separate single-wavepacket terms. The local decoherence rate is always positive, while the non-local decoherence can be of either sign, and is always negative when the wavepackets become sufficiently close.

The total decoherence rate is given by the sum of these two contributions $ \Gamma^{\rm Dec}=\Gamma^{\rm Dec}_{\rm L} + \Gamma^{\rm Dec}_{\rm NL}.$ To make the formal analogy with the superradiant emission more transparent, we shall consider the ratio of the total decoherence rate to the local decoherence rate $\Gamma^{\rm Dec}/\Gamma^{\rm Dec}_{\rm L}  = 1 + \Gamma^{\rm Dec}_{\rm NL} / \Gamma^{\rm Dec}_{\rm L}.$ Indeed, when the wave-packet distance is larger than several transition wavelengths, the quantum interference contribution to the decoherence rate becomes negligible ($\Gamma^{\rm Dec}_{\rm NL} \ll \Gamma^{\rm Dec}_{\rm L}$) and thus the decoherence rate coincides with the spontaneous emission rate. This situation is in formal analogy to the incoherent spontaneous emission of sufficiently distant atoms. In the opposite limit, when the two wavepackets are close enough, the interference contribution may reduce and eventually cancel the total decoherence rate, which is analogous to the Dicke superradiance of two emitters at subwavelength distance. The main difference is the presence of a minus sign in the non-local decoherence rate. A diminution of the center-of-mass decoherence of a single emitter is thus formally equivalent to an enhancement of the collective photonic emission.

The local and non-local phases~(\ref{eq:localnonlocal phases}) involve second-order correlation functions between the dipole and the electric field operators at different times and positions.  The  dipole correlations are expressed as  $\langle e | d_i(0) d_j(\tau)| e \rangle=\frac 1 3  \delta_{ij}  |\mathbf{d}|^2 e^{i \omega_{0} \tau} $ with $\omega_0$ the transition frequency. The local and nonlocal decoherence rates can then be recast in terms of the trace of the Green's dyadic $\mathcal{G}(\vb{r},\vb{r}';\omega_0)=  \Tr G(\vb{r},\vb{r}';\omega_0)$~\cite{NovotnyBook}:
\begin{eqnarray}\label{eq:dec_rate_I}
    \Gamma^{\rm Dec}_{\rm L} &= & \displaystyle \frac{\pi c}{\omega_0} \Gamma_0 \Im{  \mathcal{G}(\vb{r}_1,\vb{r}_1;\omega_0)+\mathcal{G}(\vb{r}_2,\vb{r}_2;\omega_0)}  \\
    \Gamma^{\rm Dec}_{\rm NL} &= & -\displaystyle \frac{2 \pi c}{\omega_0} \Gamma_0 \Im{ \mathcal{G}(\vb{r}_1,\vb{r}_2;\omega_0) } \label{eq:dec_rate_C}
\end{eqnarray}

where $\Gamma_0 = \omega_0^3 |\mathbf{d}|^2/(3 \pi \hbar \varepsilon_0 c^3)$ is the spontaneous emission rate. As a consistency check, in the absence of scattering contribution to the Green's function (i.e. $\mathcal{G}(\vb{r},\vb{r}';\omega_0) =   \mathcal{G}^{0}(\vb{r},\vb{r}';\omega_0)$), by using  Eqs.~(\ref{eq:TrG0},\ref{eq:dec_rate_I},\ref{eq:dec_rate_C}), one retrieves the free-space decoherence rate~\cite{Reinaldo16}.

\subsection{Decoherence near a material undergoing a percolation transition}

\begin{figure}[htbp]
    \centering
    \subfigure
    {\includegraphics[width=8.5cm]{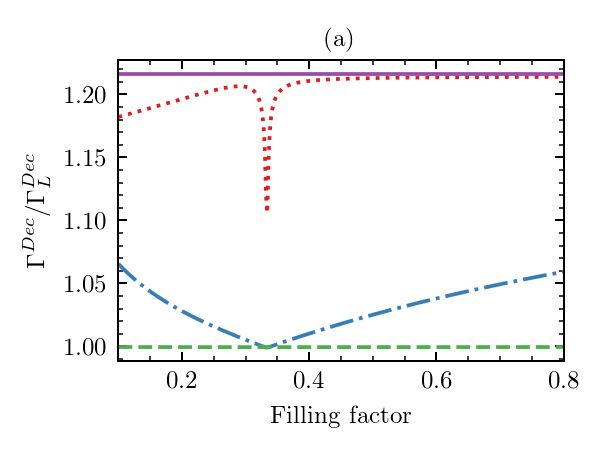}} \\
    \subfigure
    {\includegraphics[width=8.5cm]{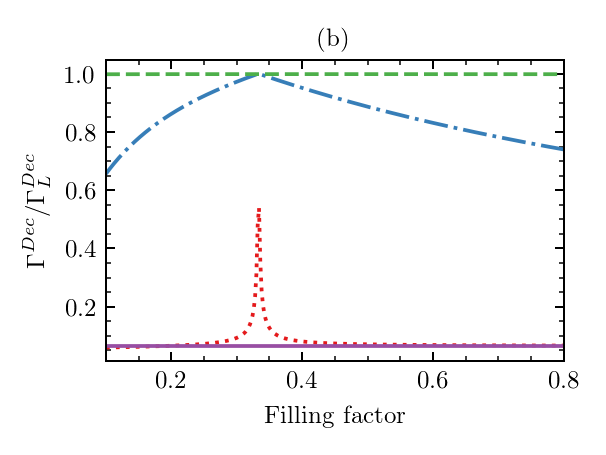}}
    \caption{\label{fig:2}(color online) Normalized decoherence rates $\Gamma^{\rm Dec} / \Gamma^{\rm Dec}_{\rm L}$ for a superposition of two wave-packets separated by a distance (a) $ x= 0.7 \lambda_0$ and (b) $ x= 0.1 \lambda_0$ as a function of the filling factor $f$ and for different distances $z$ to the composite metal-dielectric material. Here we consider $z = 10^{-2} \lambda_0$ (red dotted line), $z = 10^{-3} \lambda_0$ (blue dash-dotted line) and $z = 10^{-4} \lambda_0$ (green dashed line). The values of the decoherence rates with no material are also plotted (full purple lines). Decoherence rates have been obtained from Eqs.(\ref{eq:dec_rate_I},\ref{eq:dec_rate_C}) by considering $\lambda_0 = 450 \: {\rm \mu m}$ and the material properties described in Section~\ref{subsection:opticalproperties}.}
\end{figure}

We have studied the ratio of the decoherence rates $\Gamma^{\rm Dec}/\Gamma^{\rm Dec}_{\rm L}$ for two specific configurations corresponding to different distances between the wavepackets. The results are shown on Fig.~\ref{fig:2}. In the first configuration, the separation between the two wavepackets corresponds $ x=0.7 \lambda_0$. For this separation, the decoherence rate in the vacuum is increased by interference effects, and is roughly $20 \%$ larger than the spontaneous emission rate. In this case, each emitted photon carries a which-way information enhanced by interference effects (i.e. greater than that at infinite wave-packet separation). In the second configuration, we consider a wavepacket separation of $ x=0.1 \lambda_0$ yielding a free-space decoherence rate reduced to only $7 \%$ of the spontaneous emission rate, and thus associated to a which-way information per photon of low quality. For each wavepacket separation, we have considered various distances from the material and the filling factor $f$ across the critical region. The results, presented on Fig.~\ref{fig:2}, show that for distances extremely close to the plate $z \simeq 10^{-4} \lambda_0$ the non-local decoherence rate is suppressed for any value of the filling factor $f$. For values of the distances $z$ to the material such that $10^{-2} \lambda_0 \leq z \leq 10^{-3} \lambda_0$, one observes a striking change in the competition between the local and non-local decoherence contributions to the decoherence rate. When the filling factor $f$ parameter is far from the critical region $f \simeq1/3$, the total decoherence rate is close to  its free-space value, resulting from the interplay between the non-local and the local decoherence rates. Close to the critical value $f_0=1/3$, however, the total decoherence rate tends towards the local decoherence rate. This shows that interference effects are strongly suppressed in the vicinity of the region $f \simeq f_0$ corresponding to the phase transition. In both configurations (Figs.2a and 2b), the total decoherence rate tends towards the spontaneous emission rate, regardless of the quality of information contained in each photon. The decoherence rate, considered as a function of the filling factor $f$,  presents a slope discontinuity as the filling factor $f$ reaches the critical value $f_0=1/3$, suggesting that the percolation transition leads to a``sudden death'' of the interference contribution to the decoherence rate. Indeed, the phase transition induces an abrupt change in the interplay between the decay channels contributing to the decoherence rate. Finally, one sees that the material exerts a significant influence on the total decoherence rate over a larger range of distances as the filling factor approaches its critical value.

\subsection{Decoherence near a $\rm{VO}_2$ material undergoing a metal-insulator transition}

We investigate here the decoherence rate near $\rm{VO}_2$.  Fig.(3a) reveals that the phase transition suppresses the coherent contribution so that $\Gamma^{Dec} / \Gamma^{Dec}_{\rm L} \to 1$. Fig.(3b) sketches the normalized decoherence rate in the phase transition during the heating cycle. Here we can see that when the material undergoes a phase transition the distance where it is capable to suppress the coherent contribution increases. We also see that for longer distances the different cycles merge since the distance between the wavepackets is large enough so that the decoherence rate is insensitive to the material properties. On the other hand, for smaller distances the lines for the heating and cooling cycles merge (and these also merge later on), showing that for small distances one cannot discern the separations $x$ between wavepackets. This information is encoded in the coherent term, which is suppressed at small distances $z$.

\begin{figure}[htbp]
    \centering
    \subfigure
    {\includegraphics[width=8.5cm]{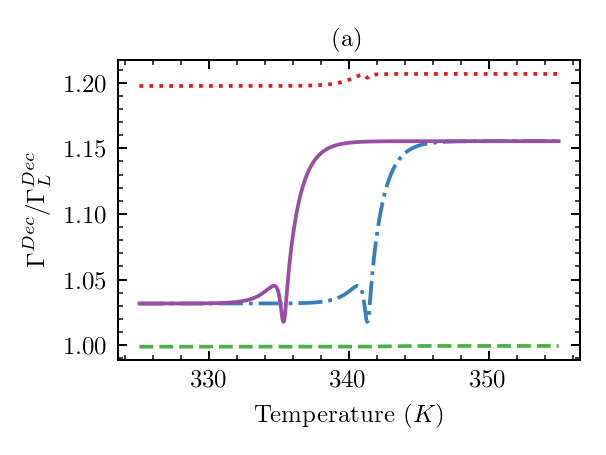}} \\
    \subfigure
    {\includegraphics[width=8.5cm]{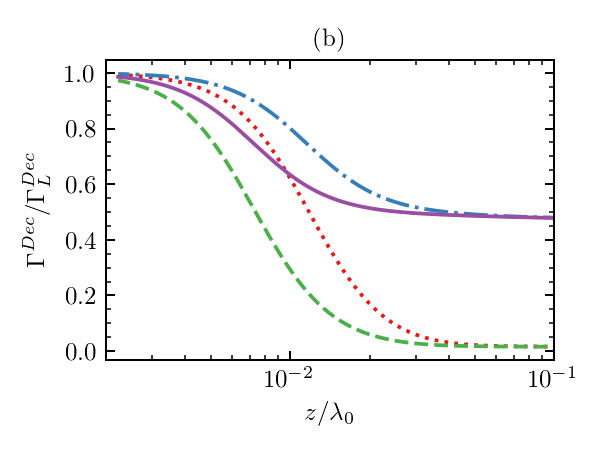}}
    \caption{\label{fig:3}(color online)   Normalized decoherence rates $\Gamma^{\rm Dec} / \Gamma^{\rm Dec}_{\rm L}$ in the vicinity of $\rm{VO}_2$ undergoing a phase transition. (a) Decoherence rate as a function of temperature $T$ for different distances $z$ to the material with a fixed separation $x = 0.7 \: \lambda_0$ between the wave-packets. We have considered for the cooling cycle $z = 10^{-2}\lambda_0$ (full purple line), and for the heating cycle  $z = 10^{-3}\lambda_0$ (green dashed line), $z = 10^{-2}\lambda_0$ (blue dash-dotted line) and $z = 10^{-1}\lambda_0$ (red dotted line). (b) Decoherence rate as a function of the distance $z$ to the material for different wave packet separations at a fixed temperature $T = 342$K. We have used $x = 0.05\lambda_0$ and $x = 0.3\lambda_0$ for both the heating cycle (red dotted and blue dash-dotted lines respectively) and the cooling cycle (green dashed and purple full lines respectively). We have taken $\lambda_0 = 450 \: {\rm \mu m}$. }
\end{figure}

\section{COLLECTIVE EMISSION NEAR CRITICAL MEDIA}

\subsection{Collective emission with prescribed dipoles}

We first recall the collective emission rate $\Gamma$ for a pair of quantum emitters, with prescribed dipole moments $\mathbf{d}_k=d_k \mathbf{u}_{k}$ oscillating in phase along fixed directions $\mathbf{u}_{k}$ in the vicinity of a semi-infinite material. This semi-classical approach is well-suited to describe superradiance from a collective symmetric quantum state of the two emitters. The global emission rate can be expressed in terms of the electric field Green's dyadic~(\ref{eq:G}) $G(\vb{r},\vb{r};\omega_0)$ evaluated at the emitter positions $\vb{r}_{m,n}$~\cite{huidobro2012}:
\begin{eqnarray}
\Gamma & = &  \sum_{m=1,2} \sum_{n=1,2}  \Gamma_{m n} \nonumber \\
\Gamma_{m  n}  & = &  \frac{2 \omega_{0}^{2} }{\hbar \epsilon_{0} c^2}  \text{Im} \left[ \mathbf{d}_{n} G (\mathbf{r}_{n},\mathbf{r}_{m}, \omega_0) \mathbf{d}_{m}  \right] 
\label{eq:totalgamma}
\end{eqnarray}
This decay rate can be written as the sum of an incoherent contribution $\Gamma_{\rm I}= \Gamma_{11}+\Gamma_{22}$ associated to the decay of two independent emitters near the surface and a coherent contribution $\Gamma_{\rm C}= \Gamma_{12}+\Gamma_{21}$ encoding interference effects in the emission process~\cite{huidobro2012}. Note that the coherent term $\Gamma_{\rm C}$ corresponds to the additional collective emission due to the simultaneous presence of several emitters. One may identify two distinct regimes, namely the regime of superradiance for which $\Gamma_C >0$ with an enhanced collective emission, and the regime of subradiance for which $\Gamma_C <0$ with an inhibited collective emission. The incoherent contribution describes the usual Purcell effect experienced by a single emitter~\cite{Szilard16}, and can be derived from the Fresnel coefficients of the material as:
\begin{equation}
\frac{\Gamma_{\rm I}}{\Gamma_{0}} =  \frac{3 c^3}{2 \omega^3_0} \text{\rm Re} \left[ \int_{0}^{\infty} dk_\parallel \frac{k_\parallel^{3}}{k_{z}} \left(1 + r^{\rm TM,TM} e^{2 i k_{z} z}  \right) \right],
\label{eq:singlegammaz}
\end{equation}
 On the other hand, the coherent contribution $\Gamma_{\rm C}$ captures the interplay between the collective effects and the Purcell effect, which may drastically change as the material undergoes a phase transition. The coherent contribution reads

\begin{equation}
\frac{\Gamma_{\rm C}}{\Gamma_{0}} =  \frac{3 c^3}{2 \omega^3_0} \text{\rm Re} \left[ \int_{0}^{\infty} dk_\parallel \frac{k_\parallel^{3}}{k_{z}} J_0(k_\parallel x) \left(1 + r^{\rm TM,TM} e^{2 i k_{z} z}  \right) \right].
\label{eq:coherentgammaz}
\end{equation}
 
We focus on this last term in order to unveil the influence of phase transitions on collective quantum emission. We investigate two different physical situations, namely the percolation metal-insulator transition in inhomogeneous media and the specific case of $\rm{VO}_2$, where the phase transition is driven by the temperature. \\

\subsection{Collective quantum emission at the percolation transition}

We now discuss the role of the percolation phase transition in the quantum emission of two quantum emitters. We consider a specific geometry where both emitters are equidistant from the material and with parallel dipoles orientations -- orthogonal to the medium interface taken as a plane.

\begin{figure}[htbp]
\begin{center}
\includegraphics[width=8.5cm]{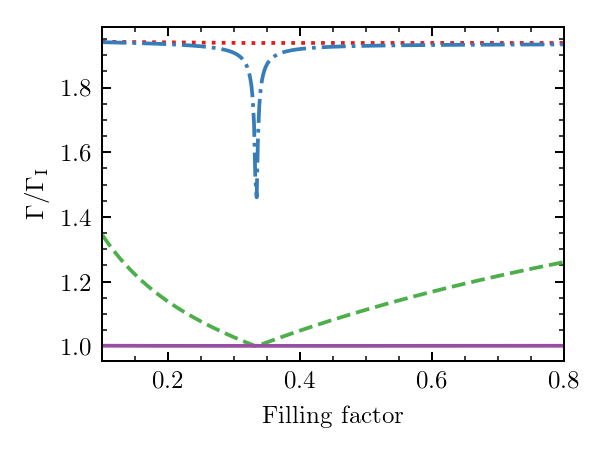}
\end{center}  \caption{(color online). Normalized spontaneous emission rate in the composite material for the symmetric state as a function of the filling factor $f$ at a separation $x = 0.1\lambda_0$ between emitters with transition wavelength $\lambda_0 = 450 \: {\rm \mu m}$. The normalized emission rate was calculated by considering eqs. (\ref{eq:singlegammaz},\ref{eq:coherentgammaz}) and the effective permittivity described in section \ref{subsection:opticalproperties}. The dotted red, blue dash-dotted, green dashed and purple full lines correspond to distances $z = 10^{-1} \lambda_0$, $z = 10^{-2} \lambda_0$, $z = 10^{-3} \lambda_0$ and $z = 10^{-4} \lambda_0$ from the material respectively.}
\label{fig:4}
\end{figure}

Fig.~\ref{fig:4} presents the normalized global decay rate $\Gamma / \Gamma_{\rm I}$ as a function of the filling fraction $f$ of gold inclusions in a polystyrene host for different values of the emitter-material distance $z$ between the emitters and the semi-infinite inhomogeneous medium. It reveals that at distances much smaller than the transition wave-length, typically of the order $z/\lambda_0 \simeq 10^{-4}$, cross-talks between the two quantum emitters are totally suppressed so that the incoherent  contribution $\Gamma_{\rm I}$ predominates  for all $f$. On the other hand, for larger distances of typically $z \gtrsim 0.1 \lambda_0$, the material properties of the compound hardly affect the emission. In this limit the emission process is entirely governed by cross-talks between the emitters. For this range of distances, one has $\Gamma \approx 2 \Gamma_{\rm I}$ for all $f$, which correspond to a maximal superradiance. Remarkably, at the critical point $f=1/3$ and for an intermediate range of emitter-material distances, there is a strong reduction of the coherent contribution to the total emission rate. In other words, collective effects are strongly suppressed at the critical point. As the collective contribution disappears with a slope discontinuity at the percolation threshold, one may speak of a ``sudden death'' of collective effects at the critical point -- as shown previously for the interference contribution to the total decoherence rate. 

\subsection{Collective emission in the vicinity of a metal-insulator $\rm{VO}_2$ transition.}

\begin{figure}[htbp]
\begin{center}
\includegraphics[width=8.5cm]{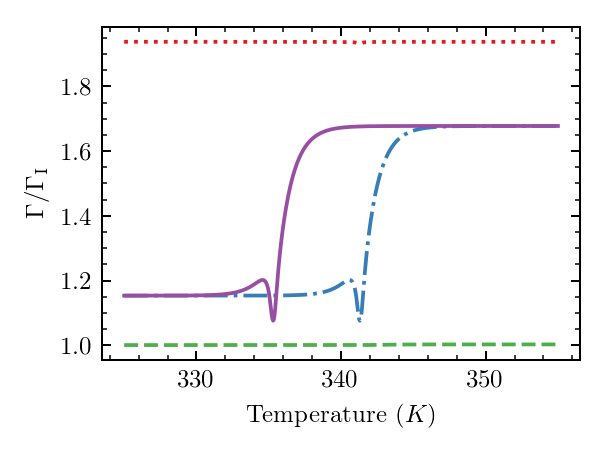}
\end{center}  \caption{(color online). Spontaneous emission rate in $\rm{VO}_2$ normalized by the incoherent contibution for the symmetric state as a function of the temperature $T$ at a separation $x = 0.1\lambda_0$ between emitters with transition wavelength $\lambda_0 = 450 \: {\rm \mu m}$. The figure shows the emission rate for the cooling cycle at $z = 10^{-2}\lambda_0$ (full purple line) and for the heating cycle at $z = 10^{-3}\lambda_0$ (green dashed line), $z = 10^{-2}\lambda_0$ (blue dash-dotted line) and $z = 10^{-1}\lambda_0$ (red dotted line).}
\label{fig:5}
\end{figure}

We now investigate the collective emission effects in the vicinity of a $\rm{VO}_2$ material undergoing a metal-insulator phase transition. We consider again the previous geometry, with   
 two quantum emitters at a fixed subwavelength separation $x = 0.1\lambda_0$ and at the same distance $z$ from the interface.  Fig.~\ref{fig:5} shows the collective emission for different values of this distance $z$. Fig.~\ref{fig:5}, where the normalized global decay rate $\Gamma / \Gamma_{\rm I}$ is calculated as a function of the temperature $T$, demonstrates that the characteristic hysteresis of $\rm{VO}_2$ also shows up on collective quantum emission. Interestingly, this implies that the total decay rate of the system will not only strongly depend on the temperature but also on whether one is cooling or heating the system. In the limit of small distances $z \leq 10^{-3} \lambda_0$, the interaction of each emitter with the medium dominates regardless of the temperature and the two emitters behave independently. On the other hand, for larger distances $z \geq 10^{-1} \lambda_0$, one has a maximal superradiance independently of the temperature. In these situations, the emitters are hardly influenced by the material. In contrast, the tuning of the collective effects by the temperature is most effective in the intermediate range $z \simeq 5 . 10^{-2} \lambda_0$. At these distances, the phase transition enhances quantum interferences between the emitters and the hysteresis is more pronounced, and thus more sensitive to the temperature. 

\section{Conclusion}

In conclusion, we have shown that the presence of a critical medium significantly alters the decoherence experienced by a single emitter, as well as the collective quantum emission. These distinct physical phenomena, namely the decoherence suffered by a single emitter in a coherent superposition of two wave-packets and the collective decay process of two quantum emitters, have been addressed within a unified framework.  Indeed, the decoherence rate and the collective emission rate are given in terms of the sum of an incoherent and a non-local contribution. The former is directly related to spontaneous emission process, whereas the latter encodes interference effects in the decoherence rate, or in the collective decay rate of two emitters.

We have found that the percolation phase transition in inhomogeneous media drastically suppresses the non-local contributions to the center-of-mass decoherence and to the collective emission by two emitters. One observes a slope discontinuity reminiscent of a ``sudden death'' of collective effects and non-local decoherence at the critical point.  In the specific case of $\rm{VO}_2$ undergoing to a metal-insulator transition driven by the temperature, the decoherence rate and collective emission exhibit a characteristic hysteresis leading to different behaviours in the cooling and heating cycles. This result suggests $\rm{VO}_2$ as an alternative material platform to tune collective effects by means of a phase transition. Altogether our results indicate that critical phenomena appear as a promising mechanism to control and tune light-matter interaction at the nanoscale.

\acknowledgements

 The authors are grateful to Paulo A. Maia-Neto, Daniela Szilard, Reinaldo F. de Melo e Souza, Marcelo Fran\c{c}a, and Felipe Rosa for stimulating discussions. This work was supported by the Brazilian agencies CNPq, CAPES, and FAPERJ. F.A.P. also thanks the The Royal Society-Newton Advanced Fellowship (Grant no. NA150208) for financial support. This work is part of INCT-IQ from CNPq.

\appendix

\section*{APPENDIX: INFLUENCE OF THERMAL EFFECTS ON THE DECOHERENCE AND COLLECTIVE EMISSION RATES}

In this appendix, we extend the results discussed in the main text to account for a possible finite temperature of the electromagnetic~(EM) field. Indeed the assumption of EM field in the vacuum state can appear as too restrictive, or even unrealistic, for emitters in the vicinity of an heated ${\rm VO}_2$ material. In this Appendix, we thus consider instead the EM field at a finite temperature $T$, described through the following thermal density matrix~\cite{MandelWolfBook}:
\begin{equation}
\label{eq:thermaldensitiy}
    \rho_{\rm EM} = \prod_k \qty(1 - \exp(\frac{\hbar \omega_k}{k_B T})) \exp(\frac{\hbar \omega_k a_k^\dagger a_k}{k_B T})
\end{equation}
where $k_B$ is the Boltzmann constant. When evaluating the electric field Green's function~(\ref{eq:dec_rate_I},\ref{eq:dec_rate_C}) with this thermal density matrix, the finite temperature merely introduces an additional factor $(n_{\omega_0}(T) + 1)$, where $n_{\omega_0}(T)$ is the average Bose-Einstein distributed photon number at the transition frequency $\omega_0$ and for the temperature~$T$. Thus, the thermal field raises a simple enhancement factor for the global decoherence rate $ \Gamma^{\rm Dec}(T) = (n_{\omega_0}(T)+1) \Gamma^{\rm Dec}(0),$ corresponding to the stimulated emission with thermal photons. Most importantly, the thermal field does not affect the decoherence profile as a function of the phase transition, nor the balance between the interference and the incoherent contributions to the total decoherence rate. The conclusions obtained for the decoherence rates with an EM field in the vacuum state thus still prevail at finite temperature.

We now investigate the influence of a finite EM field temperature on the collective emission. For this purpose, we use the field thermal density matrix~(\ref{eq:thermaldensitiy}), assuming EM field and the emitter d.o.f. initially uncorrelated, and take the trace over the EM field d.o.f. in the Born-Markov master equation~\cite{BreuerBook}:
\begin{align}\label{eq:born-markov}
    \pdv{\rho_S}{t}(t) & \! = \! \! \! \! \! \! \! \displaystyle \sum_{(m,n) \in \{1,2 \}^2} \! \! \! \! \! \! \!  (n_{\omega_0}(T)+1)\frac{\Gamma_{mn}}{2}(2 \sigma_m \rho_S \sigma_n^\dagger \nonumber \\
    & \qquad \qquad  \qquad \qquad  \qquad \qquad - \sigma_m^\dagger \sigma_n \rho_S - \rho_S \sigma_m^\dagger \sigma_n) \nonumber \\
& + n_{\omega_0} (T) \frac{\Gamma_{mn}}{2}(2 \sigma_m^\dagger \rho_S \sigma_n - \sigma_m \sigma_n^\dagger \rho_S - \rho_S \sigma_m \sigma_n^\dagger)
\end{align}
with $\rho_S$ the reduced density matrix corresponding to the emitters d.o.f., with the quantum operators $\sigma_m$ acting upon the two-level emitter $m$ as $\sigma_m | e_m \rangle = | g_m \rangle $ and $\sigma_m | g_m \rangle =0 $, and with the previously introduced dipole couplings $\Gamma_{mn}$~(\ref{eq:totalgamma}). The population $\rho_{s,s}(t)= \langle s | \rho_S(t)  | s \rangle $ in the symmetric state  $\ket{s} = \frac{1}{\sqrt{2}}(\ket{eg}+\ket{ge})$ follows the rate equation
\begin{align}\label{eq:pop_s_thermal}
    \pdv{\rho_{s,s}}{t}  = \displaystyle & (\Gamma_{\rm I} + \Gamma_{\rm C}) \left[  n_{\omega_0} (T) \rho_{gg,gg} + (n_{\omega_0}(T) + 1) \rho_{ee,ee} \right. \nonumber \\
    & \qquad \qquad \qquad \left. - (2 n_{\omega_0} (T) + 1) \rho_{s,s} \right]
  \end{align}
 $\rho_{gg,gg}$ and $\rho_{ee,ee}$ denote the density matrix population for two emitters simultaneously in the ground state and in the excited state respectively.  Fom the equation above, we infer that the decay of the symmetric state is given by the rate $(2 n_{\omega_0} (T) + 1) (\Gamma_{\rm I} + \Gamma_{\rm C}).$ The enhancement by a finite temperature factor 
  $(2 n_{\omega_0} (T) +1)$ with respect to the vacuum EM field state corresponds to stimulated emission and is on the order of $20$ for the temperatures and frequencies considered in the main text. Nevertheless, the ratio between the coherent contribution to the collective emission rate and the total emission rate remains unchanged by the finite field temperature. We conclude that taking into account thermal effects in the field itself on the collective emission does not qualitatively affect the suppression of the coherent contribution at the phase transition as well as the other findings presented in this article.


\begin{thebibliography}{1}
\bibitem{Footnote1} Note that when considering the decoherence of a single delocalized quantum emitter, we shall handle the position $\vb{r}$ as a quantum operator.
\end{thebibliography}
\end{document}